**RiD-kit: Software package designed to do enhanced sampling using reinforced dynamics**

Jiahao Fan[1,3,5], Yanze Wang[1,4,5], Dongdong Wang[1], Linfeng Zhang[1,2*]

[1]DP Technology, Beijing, China. [2]AI for Science Institute, Beijing China. [3]Institute of theoretical physics, School of Physics, Peking University, Beijing, China. [4] Present affiliation: Department of Chemistry, MIT, Cambridge, MA, USA. [5]These authors contributed equally: Jiahao Fan, Yanze Wang

*To whom correspondence should be addressed. Linfeng Zhang, DP Technology, Beijing 100080, China. AI for Science Institute, Beijing 100080, China. Email: linfeng.zhang.zlf@gmail.com

## Abstract

Developing an efficient method to accelerate the speed of molecular dynamics is a central theme in the field of molecular simulation. One category among the methods are collective-variable-based methods, which rely on predefined collective variables (CVs). The difficulty of selecting a few important CVs hinders the methods to be applied to large systems easily. Here we present a CV-based enhanced sampling method RiD-kit, which could handle a large number of CVs and perform efficient sampling. The method could be applied to various kinds of systems, including biomolecules, chemical reactions and materials. In this protocol, we guide the users through all phases of the RiD-kit workflow, from preparing the input files, setting the simulation parameters and analyzing the results. The RiD-kit workflow provides an efficient and user-friendly command line tool which could submit jobs to various kinds of platforms including the high-performance computers (HPC), cloud server and local machines.

## Introduction

Molecular dynamics (MD) have been essential tools in modeling chemical reactions, biomolecules and materials. However, the time scale problem due to the free energy barriers prevents the methods from reaching the experimental timescale for a lot of interesting cases such as protein folding[1]. To tackle the time scale problem, numerous enhanced sampling methods have been developed over the years. These methods have their roots in two major categories: **tempering methods** like replica exchange molecular dynamics (REMD)[2-4] and integrated tempering sampling(ITS)[5,6], and **collective-variable-based methods** like umbrella sampling (US)[7], metadynamics(MTD)[8,9], adaptive biasing force method (ABF)[10] and on-the-fly probability enhanced sampling (OPES)[11].

The tempering methods do not rely on predefined collective variables (CVs), but the temperature ladder increases rapidly with the degrees of freedom of the system. The CV-based methods explore the configuration space along collective variables, and they share the common feature of adding biased potentials or biased forces to the original system. Conventional CV-based methods, like metadynamics[8,9], deposit Gaussian functions along the historical trajectories in CV space, but the number of functions needed to reconstruct the potential of mean force (PMF) increases exponentially with the number of CVs. One way to solve this problem is to limit the number of collective variables, which is far from trivial. Another way is to directly fit the high-dimensional

FES by nonlinear functions. Recent advances in machine learning make high dimensional PMF reconstruction possible by gaussian process regression (GAP)[12,13] or neural networks[14-17]. However, accurately representing the PMF of a complex system requires a sufficient sampling of the conformation space. The "exploration-fitting" procedure is a chicken-egg problem, sharing a similar flavor with the "exploration-exploitation" trade-off in the context of reinforcement learning (RL)[18]. With this analogy, Reinforced Dynamics (RiD)[15,16] was developed in 2017 as an enhanced sampling scheme that can deal with high-dimensional PMF in a RL formulation. A detailed description of the RiD method development is given below.

**Development of the protocol**

RiD is a CV-based enhanced sampling method. RiD borrows the idea from reinforced learning and has the same analogy of state space, action space, reward function and the best policy. The state space is defined as the collective variable space where RiD would explore. The action space is the bias potential estimation predicted by neural networks, i.e., where to deposit bias potential. The reward function is the uncertainty of the PMF predictions so RiD is driven to explore the unknown areas in CV space. The best policy is the inverted PMF ground truth. Starting from a point in the state space, RiD explores the CV space via biased molecular simulations where the bias was the inverted predicted PMF. The uncertainty indicator, i.e., the model deviations of PMF predictions, will guide our biasing policy to only bias the areas with low uncertainty and pick configurations from high-uncertainty areas to update the PMF predictors. Due to the high dimensionality of the PMF, RiD uses an ensemble of neural network models to learn the mean forces (MF) as a proxy. Iteratively, RiD explores the CV space, generates training data on-the-fly and learns the PMF. This type of “exploration-labeling-training” procedure has been named as “concurrent learning” and also introduced elsewhere[19]. Though the concurrent learning shares much similarity with the active learning scheme, such as the usage of data uncertainty, they are fundamentally different. Active learning requires a pre-sampled dataset and aims to find the minimal subset for labeling. Concurrent learning doesn’t require any preceding data and is able to generate by itself, concurrently with the exploration procedure.

RiD is considered an application of the concurrent learning scheme on PMF. Despite the powerful framework, learning PMF is not a trivial task. Instead of learning the ill-defined values of free energy,

RiD learns PMF by fitting their derivatives, i.e., the mean force (MF). RiD offers two schemes to calculate MF labels for configurations: the restrained MD methods and the constrained MD methods. Along with the previous works[15,16], restrained MD restrains the CVs by harmonic potentials and approximates PMF as the spring force under stiff spring limits. Restrained MD is capable with most types of collective variables, such as angles, distances, RMSD or even highly nonlinear ones, but one has to tune the force constants to achieve optimal results. On the other hand, constrained molecular dynamics[20,21] algorithmically fixes CVs to the predetermined values, avoiding hyperparameter tuning. However, only simple types of CVs (distances and angles) are supported for constraints due to the algorithmical issues.

With the recent advancement of scientific workflow managers, the complicated computing and data workflow can be efficiently managed by automatic workflow agencies such as Argo[22] and Dflow[23]. The agency helps parallel molecular simulations, dispatch tasks, allocate resources, monitor task progress and decouple software environments of different computing phases. We harness the power of scientific workflow toolkits and equip RiD with the high-efficient task manager Dflow.

In this paper, we introduce RiD-kit[24], an open-source software package, for performing reinforced dynamics (RiD) and the PMF reconstruction. The RiD-kit workflow is managed by Dflow, compatible with multiple computation platforms over the local machines, high performance cluster (HPC) and cloud platform. RiD-kit is currently interfaced with GROMACS[25] and LAMMPS[26]. RiD supports conventional classical force fields as well as Deep learning potential[27].

## Applications

**RiD has been proven able to speed up simulations on various systems[15,16] across different scales. (Table 1).**

| System | Conventional methods reference | RiD time | Error in energy barrier or PMF |
|---|---|---|---|
| Ala-dipeptide[15] | 5.1 $\mu s$ MD | 20.8 ns | ~2 kJ/mol |
| Ala-tripeptide[15] | 47.7 $\mu s$ MD | 200 ns | qualitative agree in PMF |
| Polyalanine-10[15] | String methods (360ns for each path) | 589 ns | < 5 kJ/mol |
| chignolin[16] | 100 $\mu s$ MD | 3.7 $\mu s$ | qualitative agree in PMF |
| Hafnium oxide reaction (this paper) | | | |

Table 1. RiD applications.

**Comparison with other methods**

Conventional CV-based methods, like metadynamics and umbrella sampling, can only serve one or two collective variables. Their computational cost increases exponentially with the number of CVs. Many workarounds have been proposed to bypass these issues. One way is to extract fewer collective variables and reduce CV dimensions such as DeepTICA[28]. However, these collective variables are highly nonlinear and lose interpretability. Another way is to fit the high dimensional FES by neural networks (DeepVES)[17] or decomposition to subspaces (BEMetaD[29] and PBMetaD[30]). DeepVES used neural networks as variational functionals to represent high-dimensional FES. But it requires fully sampled data in advance to learn the bias potential. BEMetaD and PBMetaD assume decoupled collective variables and deposit bias potential marginally on each dimension. This procedure boosts sampling but can only get the marginal function along each dimension rather than the joint PMF representation.

RiD tackles all the problems by concurrent learning and neural network models. It doesn't require pre-sampled data as the data can be generated concurrently with the exploration step. RiD represents the joint PMF by learning MF via neural networks and always keeps the interpretability from the well-defined collective variables.

**Advantages and limitations of this protocol**

**Advantages**

**Capability to handle a large number of CVs**: Since RiD uses neural networks to fit the PMF, unlike MetaD which uses the summed Gaussian potentials, it could overcome the curse of dimensionality and handle the high dimensional data. In this manner, the requirement to select under 4 CVs is alleviated and the users could use a sufficient number of CVs which are necessary for the system dynamics.

**Concurrent learning framework**: RiD generates data on-the-fly. Data are sampled during the exploration phase, picked by uncertainty indicators and updated for model training. No need to pre-sample before simulations. This feature is in contrast with, for example, the WHAM method with umbrella sampling, or PMF reconstruction with TAMD, where the two methods first perform the sampling and then reconstruct the PMF.

**Automatic PMF calculation**: The module for calculating the potential of mean force (PMF) from the trained model is implemented in RiD-kit, minimizing the user's effort. RiD-kit uses MCMC with the trained model in the CV space, and can obtain the projected PMF in the specified dimensions.

**Computational infrastructure agnostic:** Our implementation of the RiD-kit software supports different computation resources including: local machine, high performance computing (HPC) platform and cloud server. For example, a typical local machine environment is a server with multiple GPU cards, a typical HPC platform is a supercomputing center with a job scheduling system like Slurm, a typical cloud server is like Bohrium server.

**Limitations**

**CV choice**: As a CV-based method, RiD still requires a set of user-defined CVs. Though one can randomly select, for example, all dihedral angles of a protein molecule as CVs, the neural network models have to fit. More CVs requires larger models and may slow down the inference and sampling.

**Computational cost:** RiD involves the biased MD simulations and mean force labeling. Due to the computational complexity and current hardware limitations, the neural network inference for bias potential calculations is slower than the classical force field calculations and analytical bias potential calculations (such as gaussian functions on grids). This leads to longer wall time for biased MD simulations. On the other hand, RiD has to perform restrained MD or constrained MD to calculate mean force labels, which requires additional computation resources. However, we also point out that RiD-kit enables parallelism and the computational time can be much reduced by massive computational resources, such as GPUs.

**Mean force accuracy & convergence:** The accuracy of PMF prediction is limited by the accuracy of mean force calculations. Mean force calculations for complicated CVs, such as RMSD and radius of gyration (Rg), are known to be hard to converge and require longer simulation time to get satisfying results. For restrained MD labeling, the force constants have to be tuned carefully to both meet the large force limitation and avoid numerical instability. Constrained MD labeling provides more decent MF calculations without hyperparameter tuning but only supports simple CVs (atom distance, angles, dihedral angles).

## Overview

In this protocol, we introduce the RiD-kit protocol. To show the conventional RiD procedure, we used the widely known peptide chignolin which a well-defined folding structure. To demonstrate the widespread usage scope of RiD-kit, we also briefly introduce an example of the Hafnium oxide production reaction (the background for this example can be found in the Supplementary Information). These examples demonstrated that RiD-kit can be employed in diverse situations over the biological, chemical and material science.

RiD-kit workflow needs three input attributes to start with: INPUTDIR, RiD_JSON and MACHINE_JSON. INPUTDIR specifies the directory containing all the input files needed by the RiD-kit workflow (e.g., conformation files, topology files and [optionally] force field files.). RiD_JSON defines all hyperparameters. MACHINA_JSON defines the computational resource configurations. After the preparation, the execution of RiD-kit is very simple:

```
rid submit −i INPUTDIR −c RiD_JSON −m MACHINE_JSON
```

The whole workflow contains four phases: *Exploration, Selection, Labeling and Training*. Each phase has its own parameters to set and files to prepare, which will be explained in detail below.

## Expertise needed to implement the protocol

The users should know the basic MD simulation setup with software GROMACS or LAMMPS. In case the users want to design customized CV, they should be familiar with customized CV programming in PLUMED2 package.

### Experimental design

The workflow of the RiD-kit protocol is shown in Fig. 1 and explained below.

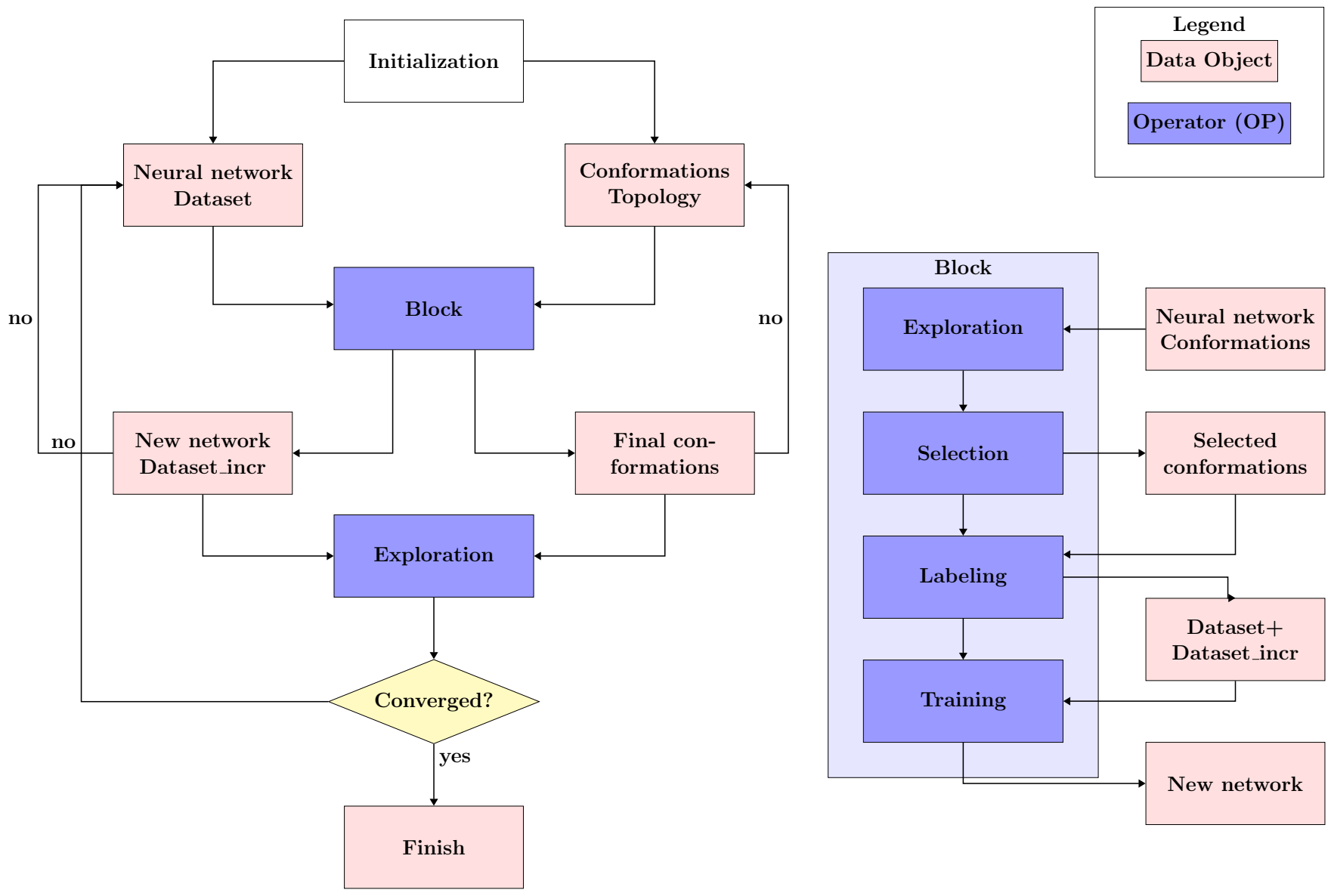

**Fig. 1** RiD-kit workflow. The main block takes neural networks and molecular topology are input. Pink boxes represent data objects while violet boxes represent operator objects as shown on top right Legend boxes.

### Overall Automatic Workflow

RiD-kit executes the exploration, selection, labeling and training steps iteratively until the PMF is converged within errors (equal to the lower bound of the trust level) or reaching the maximum iteration number. In each iteration, the simulations in the exploration phase are biased by updated PMF models from the last iteration, and newly generated configurations will be labeled and used together with the accumulated data to train the updated PMF models for the next iteration. Users are supposed to only interact with RiD-kit at the beginning and post-processing parts while RiD-kit would automatically finish the iteration loop with the workflow agency Dflow.

### Exploration

In the exploration phase, the users perform (biased) molecular dynamics simulations. The molecular configurations and topology are required for MD simulations. If the force field parameters are not involved in the topology file, the force field files are also necessary as input. RiD-kit supports both GROMACS and LAMMPS engines for simulations, so the files are expected in the corresponding formats. During simulations, the CV values and configurations are recorded. The keywords "CV" and "ExploreMDConfig" in the RiD_JSON file specifies the parameters of this stage.

**Selection**

In the selection phase, CV values generated during the Exploration phase are clustered and the values with high model deviations are collected based on the user-defined trust level. The keywords "SelectorConfig" in the RiD_JSON file specifies the values of this stage. The cluster centers are generated in the Selection, which will be sent to the Labeling phase for PMF calculations.

**Labeling**

In the labeling phase, RiD-kit calculates the mean forces for selected configurations. Currently, RiD-kit offers two labeling methods: "restrained" and "constrained", standing for restrained MD and constrained MD respectively. Simulation hyperparameters should be similar to those in the Exploration Step, such as temperatures and pressures. The simulation time for the Labeling step has to be chosen carefully to balance between convergence and cost. We recommend 100 ps for torsion mode in the restrained MD scheme, and 1 ns for distance mode in constrained MD method. For other collective variables, we recommend performing multiple long restrained simulations and confirm convergence by running averages and block analysis.

**Training**

In the training phase, an ensemble of neural networks is trained to predict mean forces from CV values. The training data are calculated from the labeling phase calculated above and accumulated every iteration. Currently, RiD-kit uses the Tensorflow framework. At least 3 models in the ensemble are recommended to provide better estimation of model deviations as the uncertainty indicator.

**Materials**

**Starting data**

1. Configuration files in .gro format (accepted by GROMACS) or in .lmp format (accepted by LAMMPS).
2. Topology file in .top format (accepted by GROMACS).

**Example data**

1. Configuration and topology files to replicate the chignolin case study, configuration file to replicate the Hafnium oxide generation case study.
2. Output files for post-processing (model files in .pdb format).
3. 2D PMF graph files in .png format.

**Hardware and software**

RiD-kit can be performed on different computational platforms including: local machine, high performance computing (HPC) platform and cloud computing server. GPU cards are highly recommended for MD simulation and model training.

The workflow of RiD-kit is managed by Dflow, where the kubernetes (also known as k8s) systems is used but not necessary. We recommend the k8s system as it provides GUI to monitor, manage and organize tasks and data. However, the deployment environment is not supposed to affect the performance.

The computational environment contains the computing software (e.g. GROMACS) and associated environment variables. On local machine or HPC platforms, the computational environment for RiD-kit should be setup manually, including:

1. Tensorflow 2.6.2, Tensorflow_cc 2.6.2, cudatoolkit packages.
2. PLUMED 2.8.1.
3. GROMACS 2022.4.
4. Other python packages, including mdtraj, nccl, cython, parmed, scikit-learn and dpdata.

The detailed installation guidance has been provided on the Github page[31].

On cloud server, the user could just specify the computation image for each phase, specifically:

1. registry.dp.tech/public/pkufjhdocker/rid-gmx-exploration:stable for the Exploration phase.
2. registry.dp.tech/public/pkufjhdocker/rid-gmx-plumed:stable for the Labeling phase.
3. registry.dp.tech/public/pkufjhdocker/rid-gmx-tf:stable for the Selection phase.

4. registry.dp.tech/public/pkufjhdocker/rid-tf-gpu:stable for the Training phase,
5. registry.dp.tech/public/pkufjhdocker/rid-tf-cpu:stable for other steps in the RiD-kit protocol.

**Experimental setup**

The running logic of the RiD-kit pipeline is contained in the RiD-kit python package and the installation of the package is very easy, just execute the following commands:

```
conda create -n rid python=3.9
conda activate rid
pip install -U rid-kit
```

This will install the "rid" entry point which controls the pipeline but will NOT install the computation software. The actual computation of each phase should be done within the computation environment specified above. The computation environment can be either docker image or conda environment based on the computation resources. For detailed instructions for installation of the computation environment, we refer the readers to the Section "Installation instructions" in Supplementary Information.

**Procedure**

As stated previously in the "overview" section, the RiD_JSON file specifies all necessary parameters for the workflow under different keyword categories. With the assistance of Dflow, RiD-kit offers one-command execution for the entire workflow all at once, but each phase (Exploration, Labeling and Training) can also be executed separately and individually with corresponding parameters.

**Overall workflow setting**

The global parameters should be set up ahead of other parameters in each phase. These parameters define the overall logic of the workflow.

```
"name": "chignolin",
"numb_walkers": 12,
"numb_iters": 30,
```

```
"trust_lvl_1": 2,
"trust_lvl_2": 3,
"init_models": [],
```

"name" specifies the user-defined name for the current workflow. "numb_walkers" means the number of parallel simulation walkers in the Exploration phase. It is 10 in our chignolin example. The following two parameters "trust_lvl_1" and "trust_lvl_2" are the lower bound and the upper bound of the trust level (unit in **kJ/mol**). The trust levels define the uncertainty criteria to apply the biased potential and select configurations, which eventually affect the accuracy and convergence of the PMF. "init_models" specifies the path for the initial free energy models if the workflow starts from some known models. Usually, it is left blank since we start from nothing.

**CV selection**

Before simulating everything, we should start with defining collective variables (CV) that can represent the essential process of the system. RiD-kit provides two commonly used CV types: "torsion" and "distance", and one "custom" type which can be defined by the users. Example of the CV selection is given in Table 2.

| CV type | Description |
|---|---|
| torsion | Torsion angles of specific residues for protein systems |
| distance | Distance between specific atoms |
| custom | Distance between specific atoms under constraints |

Table 2. CV selection examples

The CV selection is specified in the "CV" keyword in the RiD_JSON file. For the chignolin example, We choose all dihedral angles of the backbone as CVs, which are defined as follows:

```
"CV": {
    "mode": "torsion",
    "selected_resid": [1, 2, 3, 4, 5, 6, 7, 8, 9, 10],
```

```
        "angular_mask": [1, 1, 1, 1, 1, 1, 1, 1, 1, 1, 1, 1, 1, 1, 1, 1, 1, 1],
        "weights": [1, 1, 1, 1, 1, 1, 1, 1, 1, 1, 1, 1, 1, 1, 1, 1, 1, 1],
        "cv_file":[""]
    },
```

The “mode” specifies the CV type that is used in the RiD-kit protocol, the “selected_resid” specifies the residue IDs (starting form 1) containing the torsion angles, note that generally two dihedral angles $\phi$ and $\psi$ are defined for a given residue, but the first residue of one protein chain (N terminal) does not have $\phi$ and the last residue of the chain does not have $\psi$. So, the number of torsion angles should be 2n-2 for a n-residue protein. The "angular_mask" keyword indicates whether the CVs are periodic over $2\pi$. The “weights” keyword assigns weights to each CV during the clustering process in the selection phase below.

For other types of CV definition, one can check Box 1.

For the distance type CV in Table 2, an example is given below:

```
"CV": {
        "mode": "distance",
        "selected_atomid": [[2,5],[5,7]],
        "angular_mask": [0,0],
        "weights": [1,1],
        "cv_file":[""]
}
```

Different from the torsion type CV, the “selected_atomid” specifies the pairs of atomic IDs (starting form 1) containing the atomic distances. Since the distance is not periodic, the value of the “angular_mask” is 0.

For the custom type CV in Table 2, an example is given below, this example is for the Hafnium oxide reaction case, 13 distances are selected as CVs, meanwhile some system variables are confined in certain range :

```
    "CV": {
        "mode": "custom",
"selected_atomid":[[161,165],[124,156],[124,161],[156,165],[107,161],[139,161],[153,161],[121,162],[121,163],[121,164],[138,161],[154,161],[153,165]],
        "angular_mask": [0,0,0,0,0,0,0,0,0,0,0,0,0],
        "weights": [1,1,1,1,1,1,1,1,1,1,1,1,1],
        "units": "A",
        "cv_file": ["colvar", "plmpath.pdb"]
    }
```

The "colvar" file has content:

```
RESTART NO
UNITS LENGTH=A

DISTANCE ATOMS=161,165 LABEL=d1
DISTANCE ATOMS=124,156 LABEL=d2
DISTANCE ATOMS=124,161 LABEL=d3
DISTANCE ATOMS=156,165 LABEL=d4

DISTANCE ATOMS=161,162 LABEL=d5
DISTANCE ATOMS=161,163 LABEL=d6
DISTANCE ATOMS=161,164 LABEL=d7
DISTANCE ATOMS=121,161 LABEL=d8
DISTANCE ATOMS=121,153 LABEL=d9

DISTANCE ATOMS=107,161 LABEL=d10
DISTANCE ATOMS=139,161 LABEL=d11
DISTANCE ATOMS=153,161 LABEL=d12
DISTANCE ATOMS=121,162 LABEL=d13
```

```
DISTANCE ATOMS=121,163 LABEL=d14
DISTANCE ATOMS=121,164 LABEL=d15
DISTANCE ATOMS=138,161 LABEL=d16
DISTANCE ATOMS=154,161 LABEL=d17
DISTANCE ATOMS=153,165 LABEL=d18

pp1: PATH REFERENCE=plmpath.pdb TYPE=EUCLIDEAN LAMBDA=0.25
ene: ENERGY

UPPER_WALLS ARG=pp1.spath,pp1.zpath AT=7.5,-3.5 KAPPA=150,150 LABEL=uwall
LOWER_WALLS ARG=pp1.spath,pp1.zpath AT=2.5,-7.5 KAPPA=150,150 LABEL=lwall
UPPER_WALLS ARG=d5,d6,d7 AT=2.6,2.6,2.6 KAPPA=150,150,150 LABEL=uwall2
UPPER_WALLS ARG=d8,d9 AT=3.0,1.1 KAPPA=150,150 LABEL=uwall3

FLUSH STRIDE=100
PRINT STRIDE=100 ARG=d1,d2,d3,d4,d10,d11,d12,d13,d14,d15,d16,d17,d18 FILE=plm.out
```

The reference configuration is specified in the “plmpath.pdb” file, which reads like:

```
REMARK TYPE=EUCLIDEAN ARG=d1,d2,d3,d4 d1=2.38722 d2=0.99005 d3=3.99886 d4=2.08580 END
REMARK TYPE=EUCLIDEAN ARG=d1,d2,d3,d4 d1=2.64067 d2=1.20917 d3=3.73831 d4=1.86004 END
REMARK TYPE=EUCLIDEAN ARG=d1,d2,d3,d4 d1=2.90639 d2=1.50430 d3=3.47935 d4=1.65185 END
REMARK TYPE=EUCLIDEAN ARG=d1,d2,d3,d4 d1=3.18130 d2=1.83919 d3=3.22240 d4=1.46871 END
REMARK TYPE=EUCLIDEAN ARG=d1,d2,d3,d4 d1=3.46322 d2=2.19574 d3=2.96795 d4=1.32108 END
```

REMARK TYPE=EUCLIDEAN ARG=d1,d2,d3,d4 d1=3.75057 d2=2.56491 d3=2.71672 d4=1.22191 END

REMARK TYPE=EUCLIDEAN ARG=d1,d2,d3,d4 d1=4.04218 d2=2.94198 d3=2.46968 d4=1.18344 END

REMARK TYPE=EUCLIDEAN ARG=d1,d2,d3,d4 d1=4.33720 d2=3.32424 d3=2.22824 d4=1.21147 END

REMARK TYPE=EUCLIDEAN ARG=d1,d2,d3,d4 d1=4.63498 d2=3.71010 d3=1.99443 d4=1.30170

This example selects the 13 distances of two pairs of atoms as CVs, the specific CVs are specified after the "PRINT" keyword in the CV file. The customized CV files are provided in the "cv_file" keyword.

Box 1. Other types of CVs.

**CRITICAL:** The CV selection is very critical in RiD-kit protocol, directly influencing the convergence of the mean force calculation in the labeling phase below. The details are given in Box 2.

The choice of CVs influences the convergence of the mean force calculation. The mean forces acting on the defined collective variables (CVs) are defined as:

$$F(s) = \nabla_s A(s), A(s) = -\frac{1}{\beta} ln\, p(s), p(s) = \frac{1}{Z} \int e^{-\beta U(r)} \delta(s(r - s) \,\mathrm{d}r$$

where $F(s)$ is the mean forces defined on the CVs, $A(s)$ is the PMF, and $p(s)$ is the probability density of the CVs, Z is the normalizing constant. RiD-kit protocol uses restrained MD or constrained MD methods to estimate mean forces. The convergence of mean forces requires sufficient sampling of the hidden degrees of freedom that are not involved in CVs, which may face a huge sampling challenge if the hidden modes have large energy barriers.

On one hand, we would like to choose as few as pivotal CVs that can represent the reactions as well as reduce the computational cost; on the other hand, we are bothered by the sampling issues of the hidden degrees of freedom. However, by selecting some auxiliary CVs to involve some hidden modes, one may achieve convergence much faster. For example, if too few distance CVs are selected to calculate the mean forces for a system (Hafnium oxide reaction), the running average of the mean forces are to the left of Fig. 2. As shown in Fig. 2, the mean forces of the 4 CVs undergo an abrupt change after 400ps of constrained MD simulation, this indicates that the system undergoes a large conformational change after 400 ps. This means that the hidden degrees of freedom are not fully sampled (this is equivalent to say that hidden degrees of freedom have energy barriers that display several metastable states of the system). After carefully selecting 12 distance CVs which display slow modes of motion, the running averages of the mean forces converge consistently.

The users should balance the CV selection based on the prior knowledge of the system. Selecting redundant CVs doesn't cause any issue, because RiD-kit handles the joint high dimensional PMF by neural networks.

Box 2. The influence of CV selection

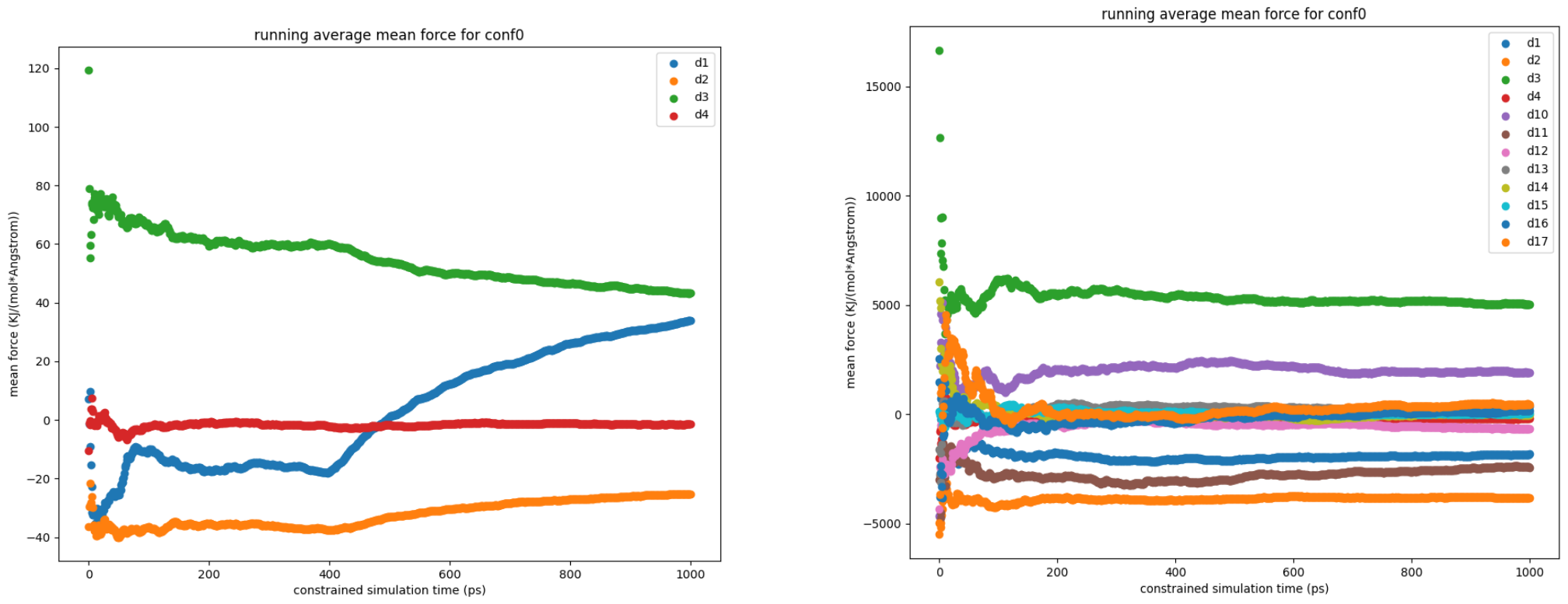

Fig. 2 Running average of the mean forces of the Hafnium oxide reaction system. The mean forces are calculated for (left) 4 distance CVs and (right) 12 distance CVs.

We assume here that GROMACS is the MD engine, which is true for the chignolin system. The reader is referred to Box 3 to learn the use of LAMMPS, which is generally used for material systems like the Hafnium oxide reaction case.

**The use of Lammps as MD engine**

The users could use LAMMPS as MD engine during the RiD-kit run, in this case, just follow the steps below:

1. Prepare the configuration files (.lmp) for the system in the input directory.

2. Prepare the input parameter files (.lammps) for LAMMPS, typically different for the Exploration and Labeling phases.

3. Set the parameters for "ExploreMDConfig" and "LabelMDConfig" in the RID_JSON file for the Exploration and Labeling phases respectively, an example config is given below:

```
"ExploreMDConfig": {
    "type":"lmp",
    "inputfile":"input_explore.lammps",
    "dt": 0.002,
    "output_freq": 50,
    "ntmpi": 1,
    "nt": 8,
    "max_warning": 2
}
```

The "type" keyword specifies the MD engine used for the Exploration phase, the "inputfile" keyword specifies the input parameter files for the LAMMPS simulation.

Box 3. The usage of LAMMPS as MD engine

**Exploration phase Timing: ~20h**

1. Parameterization (topology file and coordinate files) before exploration: the users are supposed to prepare the topology and coordinate files. The biomolecular systems, such as chignolin in this example, typically starts from a .pdb file. An example preparation procedure is given in Box 4.

1. Generate the input files for simulation (5 min): Generate the Coordinate file (.gro) and topology file (.top) as input files for the GROMACS. An example command script is given below (suppose the input .pdb structure file is ${protein}.pdb),

```
grep -v HETATM ${protein}.pdb > ${protein}_tmp.pdb
```

```
grep -v CONECT ${protein}_tmp.pdb > ${protein}_protein.pdb
gmx pdb2gmx -f ${protein}_protein.pdb -o ${protein}_processed.gro -water tip3p -ignh -ff "charmm36-mar2019"
gmx editconf -f ${protein}_processed.gro -o ${protein}_newbox.gro -c -d 1.0 -bt dodecahedron
gmx solvate -cp ${protein}_newbox.gro -cs spc216.gro -o ${protein}_solv.gro -p topol.top
```

2. Energy minimization and equilibrium run for the input system (Variable): Run the energy minimization, NVT equilibrium and NPT equilibrium,

```
gmx grompp -f ions.mdp -c ${protein}_solv.gro -p topol.top -o ions.tpr
printf "SOL\n" | gmx genion -s ions.tpr -o ${protein}_solv_ions.gro -conc 0.15 -p topol
-pname NA -nname CL -neutral
gmx grompp -f minim.mdp -c ${protein}_solv_ions.gro -p topol.top -o em.tpr
gmx mdrun -ntmpi 1 -nt 8 -v -deffnm em
gmx grompp -f nvt.mdp -c em.gro -r em.gro -p topol.top -o nvt.tpr
gmx mdrun -ntmpi 1 -nt 8 -v -deffnm nvt
gmx grompp -f npt.mdp -c nvt.gro -r nvt.gro -t nvt.cpt -p topol.top -o npt.tpr
gmx mdrun -ntmpi 1 -nt 8 -v -deffnm npt
```

3. (Optional) If using specified force field, download the force field files (5min): i.e., “charmm36-mar2019.ff” force field files.

Box 4. Preparation of topology and coordinate files

2. (Optional) If using pretrained free energy models to bias the MD simulation, put the free energy models in the input directory.
3. Set the parameters for the Exploration phase in the “ExploreMDConfig” keyword in the RiD_JSON file (10min):The parameters for MD simulations are highly case-specific due to the choice of force fields and desired simulated protocol. RiD-kit provides parametric dictionary input as below. For our example of the chignolin system, the config information for the Exploration phase is given below:

```
"ExploreMDConfig": {
        "nsteps": 2500000,
```

```
        "type": "gmx",
        "temperature": 300,
        "output_freq": 50,
        "ref-t": "300 300",
        "verlet-buffer-tolerance":"-1",
        "rlist": 1,
        "rvdw": 0.9,
        "rvdw-switch": 0,
        "rcoulomb": 0.9,
        "rcoulomb-switch": 0,
        "epsilon-r":1,
        "epsilon-rf":80,
        "dt": 0.002,
        "fourierspacing": "0.12",
        "output_mode": "single",
        "ntmpi": 1,
        "nt": 8,
        "max_warning": 2
}
```

The optimal parameters may vary depending on the force field type used, some recommendations can be found in the website[https://manual.gromacs.org/current/user-guide/force-fields.html].

5. Execute the following commands:

```
rid submit -i INPUT_DIR -c RiD_JSON -m MACHINE_JSON
```

, which will initiate the Exploration phase and the following other phases.

6. The Exploration phase records the trajectories and CVs. The CVs time series are recorded by PLUMED in “plm.out” files for each walker under the directory “plm_out”, The coordinates are recorded in “traj_comp.xtc” files for each walker under the directory “trajectory”. “confout.gro” files under the directory “conf_outs” for each walker records the final configuration. Additional files would be generated if bias potentials are applied, including “bias.png” representing the biased potential applied and the “model_devi.png” files representing the model deviation for the conformations during the simulation trajectory.

**Selection phase Timing: ~5mins**

The input files for the selection phase are automatically prepared by RiD-kit, which includes the trajectories and CV output files generated by the exploration phase. The users only need to set the parameters.

1. Set the parameters for the Selection phase in the “SelectorConfig” keyword in the RiD_JSON file (10min): Selection phase clusters the CV values and selects one representative CV in each cluster. Set the "cluster_threshold" to be the value of the initial guess of cluster threshold. Set the “numb_cluster_lower” and “numb_cluster_upper” values to be the interval of the cluster threshold. From the initial guess of the cluster, threshold will be adjusted to make the number of clusters fall into this interval. This process only happens in the first iteration. The threshold will be fixed in the following iterations. “max_selection” is the max selection number of clusters during the Selection phase. If the number of clusters from the algorithm is greater than this threshold, only the first max_selection clusters will be selected. “numb_cluster_threshold” is the value which if the number of clusters of MD trajectories in exploration step at current iteration is less than, the “trust level” of the CV will be adjusted. See published paper[16] for detail. A recommended value is half of “numb_cluster_lower”. For our example of the chignolin system, the config for the selection phase is given below:

```
"SelectorConfig": {
        "numb_cluster_lower": 30,
        "numb_cluster_upper": 45,
        "cluster_threshold": 2.6,
        "max_selection": 50,
        "numb_cluster_threshold": 15,
        "slice_mode": "gmx"
    }
```

2. After the selection phase, the selected conformation files will be generated for each walker under the directory name “selected_confs”, and the CV values for each selected conformation are recorded under the directory name “selected_cv_init”.

**Labeling phase Timing: 10-20 mins**

Labeling phase allows you to calculate mean forces for the selected configurations from the selection phase. The CV and mean force pairs are the training data for neural network models. The Labeling phase requires the selected configuration files (prepared automatically by RiD-kit in last phase) and generates the calculated mean forces.

1. Set the parameters for the labeling phase in the “LabelMDConfig” keyword in the RiD_JSON file (10min): Currently, RiD-kit supports two methods of labeling: "restrained" and "constrained", stand for restrained MD and constrained MD respectively. Most settings are quite similar to those in the Exploration phase. The simulation time is usually different between Exploration and Labeling phase, while the simulation time for Exploration has more freedom, the simulation time for Labeling phase has to be chosen with care (longer enough to ensure convergence and shorter enough to avoid wasting). Our experience is that 100ps is enough for “torsion” mode in “restrained MD” method, 1ns is enough for “distance” mode in “constrained MD” method.
   A. Restrained MD method: Set "method": "restrained" to use restrained MD as mean force calculator. The only different parameters with Exploration step are “kappas” and “std_threshold”, where “kappas” is a list of force constants ($\kappa$) of harmonic restraints where the list length is equal to the number of CVs(recommended value is 500 for the dihedral CV, the unit is kj/mol), “std_threshold” (default 5.0, the unit is consistent with the mean force) is a number representing the mean force standard deviation threshold, beyond which the mean force is neglected and will not be used in the dataset for training free energy model. You should test labeling MD for your own system to determine an appropriate number for this threshold.
   B. Constrained MD method: Set "method": "constrained" to use constrained MD as mean force calculator. Currently RiD-kit only supports distance CV to use this method, also only “gmx” type is supported to perform constrained MD simulation. The other parameters are the same with the exploration phase.

**CRITICAL**: Note that if you want to use constrained MD as the mean force calculator, apart from setting method to be constrained in the" LabelMDConfig", you should add "[ constraints ]" line corresponding to the "[ moleculartype ]" in the input topology file yourself, since GROMACS specifies constraints information for each "[ moleculartype ]".

The config of our example chignolin system for the labeling phase is given below, which uses the restrained method for performing restrained molecular dynamics:

```
    "LabelMDConfig": {
        "nsteps": 50000,
        "temperature":300,
        "method": "restrained",
        "type": "gmx",
        "output_freq": 100,
        "ref-t": "300 300",
        "rlist": 1,
        "verlet-buffer-tolerance":"-1",
        "rvdw": 0.9,
        "rvdw-switch": 0,
        "rcoulomb": 0.9,
        "rcoulomb-switch": 0,
        "epsilon-r":1,
        "epsilon-rf":80,
        "dt": 0.002,
        "fourierspacing": "0.12",
        "output_mode": "single",
        "ntmpi": 1,
        "nt": 8,
        "max_warning": 2,
        "kappas": [ 500, 500, 500, 500, 500, 500, 500, 500, 500, 500],
        "std_threshold": 1.5
```

```
    }
```

For the usage of the constrained MD method one can check Box 5.

**The usage of constrained MD method**

The advantage of constrained MD method over restrained MD method is that it does not depend on external hyperparameters like “kappas” in the restrained MD method, thus providing a much more efficient way to calculate the mean force for a given system. But in the current implementation of the rid-kit, only the distance CV supports the constrained MD method, so for other types of CV, one can only use the restrained MD method in the Labeling phase. We give one example of the config for the constrained MD method, which is used for Hafnium oxide reaction case :

```
"LabelMDConfig": {
        "type": "gmx",
        "method": "constrained",
        "temperature": 372,
        "dp_model": ["graph-compress.003.pb"],
        "nsteps": 1000,
        "ref-t": "372 0",
        "gen-vel": "yes",
        "gen-temp": "372",
        "tc-grps": "top bottom",
        "dt": 0.001,
        "tau-t": "0.05 0.05",
        "constraints": "none",
        "comm-grps": "top",
        "output_freq": 10,
        "output_mode": "both",
        "freezegrps": "bottom",
        "freezedim": "Y Y Y",
```

```
        "pcoupl": "no",
        "ntmpi": 1,
        "coulombtype": "cutoff",
        "rlist": 1,
        "rvdw": 0.7,
        "rvdw-switch": 0,
        "rcoulomb": 0.7,
        "rcoulomb-switch": 0,
        "nstenergy": 1000,
        "nt": 8,
        "max_warning": 3
    },
```

Box 5. The usage of constrained MD method

2. After the labeling phase, "cv_forces.out" files are produced for each starting configuration in the directory "cv_forces", which represents the CV values and the mean forces estimations for each configuration.

**Training phase Timing: ~5mins**

The input files for the training phase, i.e. the training data, are automatically prepared by the RiD-kit protocol from the labeling phase.

1. Set the parameters for the Exploration phase in the "Train" keyword in the RiD_JSON file (10min): "numb_models" parameter is the number of models that are trained in the Training phase. RiD-kit uses model deviations, i.e. the standard deviation of outputs of these models, to evaluate the quality of free energy surface, so "numb_models" must be greater than 1. "neurons" is a list of integers which represents the number of neurons of each layer. RiD-kit uses MLP as the basic neural network structure. Number of elements in a list means the number of hidden layers and each element defines the number of nodes in each layer. For example, [ 50, 50, 50, 50 ] means there are 4 hidden layers and each hidden layer has 50 neurons. "resnet" is a Boolean variable which decides whether to use residual connection between layers. If true, the number of nodes of layers must be equal. “epoches” is the number of training epochs. “init_lr” is the initial learning rate which will decay exponentially during training. “decay_steps” is the number of decay steps of learning rate. “decay_rate” is the decay rate of learning rate. “drop_out_rate” is the dropout rate of dropout layers. “numb_threads” is the number of threads of training. The config used for the chignolin example for the training phase is given below:

```
"Train": {
    "numb_models": 4,
    "neurons": [ 300, 300, 300, 300 ],
    "resnet": true,
    "batch_size": 128,
    "epoches": 16000,
    "init_lr": 0.0008,
    "decay_steps": 120,
    "decay_rate": 0.96,
    "drop_out_rate": 0.1,
"numb_threads": 8
}
```

2. After the training phase is completed, there will be trained models in .pb format generated under the directory “model” whose number equals “numb_models” in the config, also figures representing the loss curve will be generated under the directory “train_fig”.

**Dimension reduction phase Timing: ~2h**

This phase is to use MCMC to reduce the dimension of the PMF represented by the trained model. The users need to provide the trained model during the RiD simulation cycle.

1. Prepare the input files (5 min): Prepare the trained models (.pb format) in a directory.

Set the parameters for the dimension reduction phase in the json file (10min): "init_models" is the list of trained free energy models generated in the RiD run. "MCMC_Config" is the detailed config data specified by the users, in which the following keywords are specified: "cv_dimension" is an integer to represent the number of CVs used in RiD. "numb_steps" is an integer to represent the number of steps for MCMC run. "numb_walkers" is the number of parallel walkers used in the MCMC run. “temperature” is the temperature (in K) used in the RiD run and MCMC simulation. "cv_upper_bound" and “cv_lower_bound” (Optional) are the list of integers which represents the upper bound and lower bound of the initial value in MCMC for each dimension of CVs, this is usually set for distance CVs. "cv_type" is the CV type for projection, supporting “dih” and “dis”, representing dihedral CVs and distance CVs. "bins" is the number of bins for projecting CVs at each dimension. "proj_mode" is the mode for projecting 2D CV, supporting “cv” and “path” mode. “cv” mode represents the common projection on the selected CV index, “path” mode represents the projection on the user defined path CV. If specifying “cv” mode as the projection mode, "proj_cv_index" is the list of projected indices for the 2D CV. Note that 1D projection is done for all CVs. If specifying “path” mode as the projection mode, there are additional parameters to set: "proj_cv_index" is the set of CV index used to defined the path CV, "path_list" is the list of CVs to defined the path, and "path_lm" is the “$\lambda$” in the path CV definition.The config of the chignolin system for the dimension reduction phase is given below:

```
{
        "name": "RiD-mcmc",
        "init_models":
["model_000.pb","model_001.pb","model_002.pb","model_003.pb"],
        "MCMC_Config": {
            "cv_dimension": 2,
            "numb_steps": 1000000,
            "numb_walkers": 2000,
            "temperature": 300,
            "proj_info": {
                "proj_mode": "cv",
                "proj_cv_index": [0,1]
            },
            "cv_type": "dih",
            "bins": 101
```

```
    }
}
```

Note that this config projects the free energy onto the CV index 0 and 1, if the users want to get the projected figure onto other CV dimensions, just modify the list values in the “proj_cv_index” keyword.

For CV types other than the dihedrals, one can check Box 6.

**Setup dimension reduction parameters for other CV types**

For CV types other than the dihedrals, we currently support the distance CV type, an example is given below, note that due to the aperiodic nature of the distance CV, the sampling range is better to be defined for a faster convergence in the keywords "cv_upper_bound" and "cv_lower_bound":

```
{
    "name": "RiD-mcmc",
    "init_models": ["model_000.pb","model_001.pb","model_002.pb","model_003.pb"],

    "MCMC_Config": {
        "cv_dimension": 20,
        "numb_steps": 100000,
        "numb_walkers": 2000,
        "cv_upper_bound":
[4.5,5.0,6.0,4.2,4.0,4.0,3.5,4.0,4.0,3.5,3.2,3.0,3.5,3.5,4.0,4.5,4.5,3.75,3.8,4.0],
        "cv_lower_bound":
[1.0,2.0,3.5,2.8,2.5,2.0,1.2,1.5,1.5,1.0,1.8,0.5,1.0,1.0,1.5,1.5,1.5,1.75,2.2,2.6],
        "proj_info": {
            "proj_mode": "cv",
            "proj_cv_index": [0,1]
        },
        "cv_type": "dis",
```

"bins": 101

}

}

If the users want to project the free energy onto the path CV rather than the original CV index, the "proj_mode" keyword can be set to "path", and "path_cv_index", "path_list" and "path_lm" keywords can be defined accordingly, an example is given below, which is used for the Hafnium oxide reaction case:

```
{
	"name": "RiD-mcmc",
	"init_models": ["model_000.pb","model_001.pb","model_002.pb","model_003.pb"],

	"MCMC_Config": {
		"cv_dimension": 13,
		"numb_steps": 100000,
		"numb_walkers": 2000,
		"cv_upper_bound":
[5.33507,4.43402,4.59772,3.40287,4.67268,5.12813,3.28143,3.74967,3.58144,5.34143,7.1301
4,4.58024,7.41678],
		"cv_lower_bound":
[2.38722,0.99005,3.99886,2.08580,3.81799,3.93331,2.768175,3.145618,3.161886,4.661101,5.
286805,4.076754,2.893047],
		"proj_info": {
			"proj_mode": "path",
			"proj_cv_index": [0,1,2,3],
			"path_list":
[[2.38722,0.99005,3.99886,2.08580],[2.64067,1.20917,3.73831,1.86004],[2.90639,1.50430,3.4
7935,1.65185],


[3.18130,1.83919,3.22240,1.46871],[3.46322,2.19574,2.96795,1.32108],[3.75057,2.56491,2.71
```

```
672,1.22191],

[4.04218,2.94198,2.46968,1.18344],[4.33720,3.32424,2.22824,1.21147],[4.63498,3.71010,1.99443,1.30170]],
                "path_lm": 0.25
            },
            "cv_type": "dis",
            "bins": 101
        }

}
```

Box 6. Setup dimension reduction parameters for other CV types

2. Execute the following commands:

```
rid redim -i INPUT_DIR -c RiD_JSON -m MACHINE_JSON
```

, which will perform MCMC in the CV space and obtain the projected PMF figure.

**Running RiD phases individually**

Many phases in the RiD-kit workflow can be run as individual workflows, for example if the users want to run the Exploration phase individually with the input configuration and topology files, the following command can be executed:

```
rid explore -i INPUT_DIR -c RID_JSON -m MACHINE_JSON
```

This command will perform unbiased or biased MD for the input configuration files.

Similarly, the Labeling phase can be run by executing the following command:

```
rid label -i INPUT_DIR -c RID_JSON -m MACHINE_JSON
```

This command will calculate the mean forces corresponding to the input configuration files.

In case the users want to run the Training phase individually (want to train the free energy model with the calculated mean forces data), the users will have to prepare the input files themselves, this involves preparing the mean forces data in the .npy format, and then execute the following command:

```
rid train -i INPUT_DIR -c RID_JSON -m MACHINE_JSON
```

This command will perform neural network training with the mean forces data prepared, and obtain the .pb files as the trained models.

**Running RiD with Deep Potential model**

The RiD-kit workflow supports the usage of the pretrained Deep Potential[27] model as MD force field. We use the Deep Potential model in the Hafnium oxide reaction example, so this will be used for our demonstration. To use it, follow the steps below:

1. Prepare the Deep Potential models (.pb files) in the input directory used for the RiD-kit workflow.

2. (Optional)If using Gromacs as the MD engine, the users will have to provide the input.json file to tell Gromacs to use the Deep Potential model, the format of the input.json file is like:

```
{
    "graph_file": "/path/to/graph.pb",
    "type_file": "type.raw",
    "index_file": "index.raw"
}
```

The graph_file means the path of the trained Deep Potential model files, the type_file is the file to specify the atom types, the index_file is the file containing indices of the atoms which should be consistent with the indices' order in .gro file but starting from zero. For the Hafnium oxide reaction case, the type_file contains a list of numbers with 0 representing the silicon atoms, 1 the oxygen atoms, 2 the hydrogen atoms, 3 the Hafnium atoms and 4 the chlorine atoms. The index_file for this case contains a list of numbers from 0 to 164, representing 165 atoms of the system.

3. Add the Deep Potential model names in the configs of the Exploration and Labeling phase, for the Hafnium oxide reaction case the updated config for the Exploration phase is like:

```
    "ExploreMDConfig": {
```

```
        "type": "gmx",
        "temperature": 372,
        "dp_model": ["graph.pb"],
        "nsteps": 500000,
        "ref-t": "372 0",
        "gen-vel": "yes",
        "gen-temp": "372",
        "tc-grps": "top bottom",
        "dt": 0.001,
        "tau-t": "0.05 0.05",
        "constraints": "none",
        "comm-grps": "top",
        "output_freq": 500,
        "output_mode": "both",
        "freezegrps": "bottom",
        "freezedim": "Y Y Y",
        "pcoupl": "no",
        "ntmpi": 1,
        "coulombtype": "cutoff",
        "rlist": 1,
        "rvdw": 0.7,
        "rvdw-switch": 0,
        "rcoulomb": 0.7,
        "rcoulomb-switch": 0,
        "nstenergy": 100,
        "nt": 8,
        "max_warning": 3
    }
```

Note that for the Hafnium oxide reaction case, the bottom Si layer is fixed, so this part of the atoms is kept to be at zero temperature.

**Troubleshooting**

Troubleshooting advice is provided in Table 3

| Procedure step | Problem | Possible reason | Solution |
|---|---|---|---|
| Exploration phase | The exploration region is not extended to other regions | 1.The PMF is not converged for the explored region.<br>2. The trust level is set too low. | 1.More configurations should be added in the Labeling phase.<br>2. Increase the trust level to reasonable values to encourage exploration. |
| Labeling phase | Calculation of mean forces does not converge | 1.CVs are not selected correctly.<br>2. The simulation length is not long enough. | 1.Users should include other CVs which are changing during the restrained MD.<br>2. Increase the length of the Labeling MD run. |
| Dimension reduction phase | The projected PMF does not converge | 1. The MCMC simulation length is not long enough.<br>2. The CV range is not set for nonperiodic CV, for example distance CV. | 1.Increase the length of the MCMC simulation.<br>2. Set the approximate CV value range for nonperiodic CVs. |

Table 3. RiD-kit Troubleshooting.

These troubleshootings are for the scientific problems encountered during the RiD-kit protocol, the troubleshootings for the technical issues during the usage of the RiD-kit are provided in the Supplementary information.

**Anticipated results**

**Chignolin case.**

The PMF projections onto the 2D dihedral CV are depicted in Fig. 3.

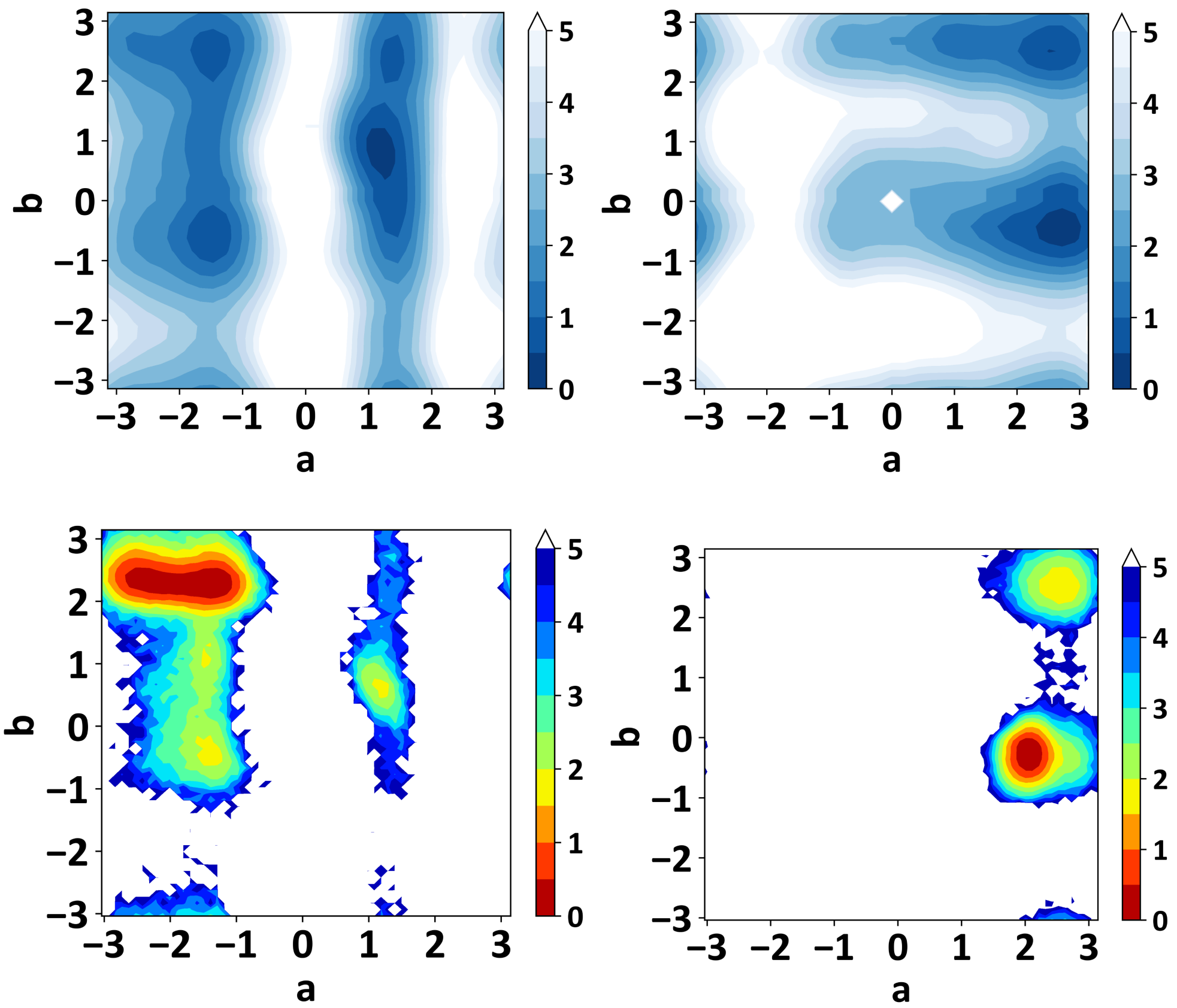

Fig. 3 Free energy projection for chignolin. (upper left and right): The PMF projections from the D.E. Shaw trajectory for the chignolin. (lower left and lower right): The PMF projections from the RiD-kit run for chignolin. (Unit: kcal/mol)

The projection is done through the dimension reduction phase of RiD-kit, specifically onto the 4th and 6th dihedral angles of the chignolin system. From Fig. 3 we can see that the PMF projections from RiD-kit fit the PMF projections from long time MD trajectory quantitatively.

**Hafnium oxide case.**

The PMF projection onto the 2D path space is depicted in Fig. 4.

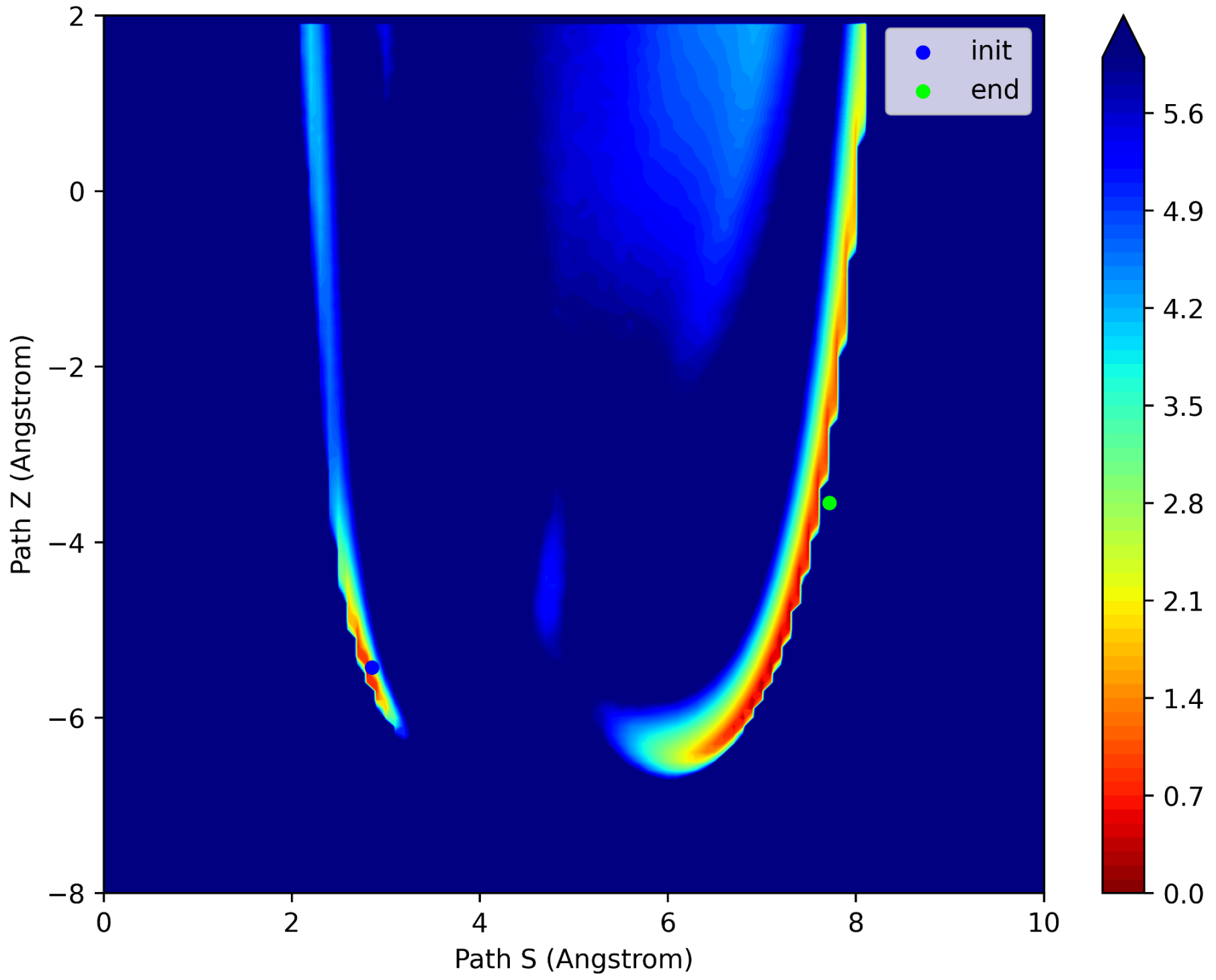

Fig. 4 Free energy projection for Hafnium oxide case. The initial configuration is depicted as a blue dot in the plot, while the final configuration through the reaction is depicted as a yellow dot in the plot. (Unit: kcal/mol)

The projection is done through the dimension reduction phase of RiD-kit, specifically onto the 2D path CV of the Hafnium oxide reaction. From Fig. 4 we can see the PMF landscape of this system.

## Supplementary information

### Background of Hafnium oxide production

In this paper, we focused on the initial stage of the Hafnium oxide ($HfO_2$) film deposition process via Atomic Layer Deposition (ALD), concentrating on the primary reaction between hafnium chloride ($HfCl_4$) and the hydroxyl groups on the silicon substrate. This foundational reaction can be represented as $HfCl_4$ reacting with Si-OH to form $HfCl_3$, hydrochloric acid (HCl), and a newly formed Si-O bond:

$$\mathrm{HfCl_4} + \mathrm{Si-OH} \rightarrow \mathrm{HfCl_3} + \mathrm{HCl} + \mathrm{Si-O}$$

Specifically, we want to calculate the reaction free energy of this process using RiD-kit. In order to do this, we need to first select collective variables (CVs) which can represent the reaction process. Since this reaction involves the breaking and formation of two chemical bonds, so an initial thought is to choose four distances between involved atoms as CVs: Hf-Cl, O-H, Hf-O and Cl-H.

Now a straightforward calculation of the free energy using metadynamics already faces problem for 4 CVs due to slow convergence, besides further investigation shows that 4 CVs may not be enough for the free energy characterization. The reason for this is that the constrained MD simulation aimed at obtaining the mean force at specific CV point is not stable during simulation process, namely many configurations will undergo a structural transition when the 4 CVs are constrained to a specific value, the phenomenon is already seen in Fig. 2 in the main text.

This is a typical case in general chemical reaction scenario that there are multiple degrees of freedom involved in the reaction process, in this case we find 13 CVs to be a good indicator. Now the number of CVs are well beyond what metadynamics could handle, so we will elaborate the advantage of RiD-kit through this example.

Since the classical potential function are lacking for this system, typically we would need to do ab-initio molecular dynamics (AIMD), but this method is very expensive and time consuming, so we choose to use machine-learning potential (ML potential) to do the MD simulation, specifically we use the well-developed ML potential method Deep Potential model[27].

After getting the ML potential, we start the RiD cycle from the initial configuration of the system. The free energy model is updated during the RiD cycle iteratively until a predefined iteration threshold is reached.

**Installation instructions**

Unless specified, all the commands in this section are executed in the Linux terminal.

**Use RiD-kit with Bohrium k8s**

The Dflow team provides a community version of k8s *deepmodeling k8s*[30], making the use of RiD-kit very convenient. To use the community version of k8s, one first needs to register a Bohrium account in Bohrium and learn a few concepts (job, jobgroup, project id) in the Bohrium website documents. Then the use of RiD-kit is very easy.

1. Set the environment variables: Just set the environment variables based on your personal Bohrium account information by the following commands:

```
export DFLOW_HOST=https://workflows.deepmodeling.com
```

```
export DFLOW_K8S_API_SERVER=https://workflows.deepmodeling.com
export DFLOW_S3_REPO_KEY=oss-bohrium
export DFLOW_S3_STORAGE_CLIENT=dflow.plugins.bohrium.TiefblueClient
export BOHRIUM_USERNAME="<bohrium-email>"
export BOHRIUM_PASSWORD="<bohrium-password>"
export BOHRIUM_PROJECT_ID="<bohrium-project-id>"
```

A convenient way is to put these lines in a bash script such as env.sh, and then execute command *source env.sh*.

**Use RiD-kit with local k8s environment**

In this section, we will learn how to deploy RiD-kit with your own k8s environment.

1. Install *minikube* as the k8s environment on your computer: See the webpage[31] for the installation detail.
2. Start *minikube* environment:

```
#choose the location of the minikube environment
export MINIKUBE_HOME=~/.minikube
#allocate enough memory based on your machine in case of high parallelism
minikube start --cpus 8 --memory 8192mb --kubernetes-version=1.23.9 --image-mirror-country='cn'
#mount the storage path if you are using machine with shared memory storage system, change the minio host path accordingly
minikube start --cpus 8 --memory 8192mb --kubernetes-version=1.23.9 --mount --mount-string="/path_on_your_machine:/data2" --image-mirror-country='cn'
```

3. Deploy *Argo* environment:

```
kubectl create ns argo
kubectl apply -n argo -f https://raw.githubusercontent.com/deepmodeling/dflow/master/manifests/quick-start-postgres.yaml
```

If you want to mount your local path into the minio path, change the minio host path in the yaml file:

```
- hostPath:
    path: /path/to/minio
    type: DirectoryOrCreate
```

4. Install RiD-kit: Install the latest RiD-kit by the following commands:

```
pip install setuptools_scm
pip install -U RiD-kit
```

**Use RiD-kit without k8s environment**

To run the workflow without k8s environment, one can use the Debug mode of Dflow. In this mode however, one cannot monitor the workflow in the *Argo* UI. Also, you need to configure the computational environment (typically conda environment) on your local machine or HPC.

1. Setup conda environment with tensorflow and tensorflow_cc packages using the commands:

```
conda create -n RiD python=3.9 libtensorflow_cc=2.6.2=*cuda110* tensorflow=2.6.2=*cuda110* -c conda-forge
```

2. Install other necessary python packages:

```
conda activate RiD
conda install mdtraj nccl -c conda-forge
conda install cudatoolkit-dev -c conda-forge
pip install cmake cython
pip install matplotlib parmed scikit-learn dpdata
```

3. Set the library path:

```
conda activate RiD
export LIBRARY_PATH=${CONDA_PREFIX}/lib:$LIBRARY_PATH
export LD_LIBRARY_PATH=${CONDA_PREFIX}/lib:$LD_LIBRARY_PATH
```

4. Install plumed 2.8.1:

```
tar -xvzf plumed-2.8.1.tgz
cp DeePFE.cpp plumed-2.8.1/src/bias
cd plumed-2.8.1
./configure --prefix=$CONDA_PREFIX \
```

```
    CXXFLAGS="-O3 -I$CONDA_PREFIX/include/" \
    LDFLAGS=" -L$CONDA_PREFIX/lib -ltensorflow_cc -ltensorflow_framework"
make -j 6 && make install
```

5. Install GROMACS 2022.4:

```
tar -xzvf GROMACS-2022.4.tar.gz
cd GROMACS-2022.4
plumed patch -p # make sure you use the correct GROMACS version
mkdir build
cd build
# if the network is on
cmake .. -DGMX_BUILD_OWN_FFTW=ON \
       -DCMAKE_INSTALL_PREFIX=$CONDA_PREFIX \
       -DGMX_GPU=CUDA
make -j 4 && make install
```

6. If using LAMMPS as MD engine, install Lammps-stable_23Jun2022_update1:

```
tar -xzvf stable_23Jun2022_update1.tar.gz
cd lammps-stable_23Jun2022_update1/src
make yes-extra-fix && make yes-extra-dump && make yes-kspace
make lib-plumed args="-p $CONDA_PREFIX -m runtime"
make yes-plumed
make serial -j4 #build serial version of lammps
ln -s /path/to/lammps-stable_23Jun2022_update1/src/lmp_serial $CONDA_PREFIX/bin/lmp_serial
```

7. Install RiD-kit: Install the latest RiD-kit by the following commands:

```
pip install setuptools_scm
pip install -U RiD-kit
```

**Troubleshooting for technical issues**

Troubleshooting advice for technical issues is provided in Table 4

| Procedure step | Problem | Possible reason | Solution |
|---|---|---|---|
| Installation | undefined reference to `std::__throw_bad_array\ _new_length()@glibcxx_3.4.29' | The version of GLICBCXX is lower than 3.4.29. | Upgrade the libstdc++6 library |
| Submission (local k8s) | Failed to establish a new connection: [Errno 111] Connection refused')': /my-bucket?location= | Port forwarding is not done by users. | Execute command: "RiD port-forward" |
| Submission (Bohrium k8s) | Failed to establish a new connection: [Errno 111] Connection refused')': /my-bucket?location= | Environmental variables are not set by users. | Set the environmental variables based on the commands in "Installation" section. |
| Exploration phase | "…/run_exploration.py", line 199, in execute "assert return_code == 0, err", AssertionError | Possible reason: network problem | Try to change the computation node |
| Labeling phase | "…/run_label.py", line 166, in execute "assert return_code == 0, err", AssertionError | Possible reason: network problem | Try to change the computation node |

Table 4. RiD-kit troubleshooting for technical issues

**Data Availability**

The RiD-kit software is publicly available at Github[24] under the LGPL-3.0 license.

**Acknowledgments**

We acknowledge computational resources provided by DP Technology.

**Author Contributions**

L.Z. and D.W. conceived the project. J.F and Y.W. implemented the workflow of RiD-kit. J.F., Y.W. and D.W. applied the RiD-kit on different systems. J.F. and Y.W wrote the manuscript.

**Declaration of interests**

The authors declare no competing interests.